\documentclass[prb,twocolumn,showpacs,preprintnumbers,amsmath,amssymb,floats]{revtex4-1}
\usepackage{epsfig}
\usepackage{setspace}
\usepackage{graphicx}
\usepackage{amsmath}
\usepackage{epstopdf}
\usepackage{amsbsy}
\usepackage{dcolumn}
\usepackage{bm}

\usepackage[usenames]{color}

\begin{document}

\title{Impurity-induced sub-gap bound gap states in alkali doped iron chalcogenide superconductors}
\author{Shantanu Mukherjee$^{1,2}$, Maria N. Gastiasoro$^2$, Brian M. Andersen$^2$}
\affiliation{
$^1$Niels Bohr Institute, University of Copenhagen, DK-2100 Copenhagen \O, Denmark\\
$^2$Niels Bohr International Academy, Niels Bohr Institute, University of Copenhagen, Blegdamsvej 17, DK-2100 Copenhagen \O, Denmark\\
}


\begin{abstract}

Measurements of the local density of states near impurities can be useful for identifying the superconducting gap structure in alkali doped iron chalcogenide superconductors K$_x$Fe$_{2-y}$Se$_2$. Here, we study the effects of nonmagnetic and magnetic impurities within a nearest neighbor $d$-wave and next-nearest neighbor $s$-wave superconducting state. For both repulsive and attractive nonmagnetic impurities, it is shown that sub-gap bound states exist only for $d$-wave superconductors with the positions of these bound states depending rather sensitively on the electron doping level. Further, for such disorder Coulomb interactions can lead to local impurity-induced magnetism in the case of $d$-wave superconductivity. For magnetic impurities, both $s$-wave and $d$-wave superconducting states support sub-gap bound states. The above results can be explained by a simple analytic model that provides a semi-quantitative understanding of the variation of the impurity bound states energies as a function of impurity potential and chemical doping level.

\end{abstract}

\pacs{74.70.Xa, 74.20.Rp, 74.55.+v, 74.81.-g}

\maketitle
\section{Introduction}
Alkali doped iron chalcogenide superconductors A$_x$Fe$_{2-y}$Se$_2$ undergo a transition to an iron vacancy ordered structure at $T_S\sim 578K$, followed by magnetic transition to a block antiferromagnetic state (BAFM) at $T_N\sim 559K$.\cite{Bao:2011,Pomjakushin:2011,Ye:2011} Below a temperature of $T_c\sim 32K$ the compounds become superconducting.\cite{Guo:2010} It is now well documented that superconductivity phase separates from the BAFM and exists in filamentary regions parallel and perpendicular to the FeSe planes.\cite{Chen:2011,Lazarevic:2012,Yuan:2012,Ding:2012,Li(a):2012,Li(b):2012,Yan:2012,Landsgesell:2012,Charnukha:2012,Ricci:2011,Liu:2012}  In addition to the presence of phase separation, any theoretical description of the superconducting state is necessarily altered compared to the usual iron pnictides by the absence of Fermi surface hole pockets.Experiments find only electron pockets around the M points of 1 Fe Brillouin zone at $k_z=0$ and the development of additional electron pockets around the Z point. \cite{Wang:2011, Zhang:2011,Qian:2011} The description of a spin-fluctuation based pairing mechanism in iron based pnictide superconductors argues for a leading $s\pm$ pairing instability based on $(\pi,0)$ nesting between the electron and hole Fermi surface sheets.\cite{Mazin:2008} The alkali doped iron chalcogenides therefore violate such arguments since the absence of hole pockets imply that the relevant nesting vector for these systems is not $(\pi,0)$ but the weaker $(\pi,\pi)$ nesting between electron pockets at the M points. Theoretical calculations for the superconducting ground state are currently inconclusive and predictions have been made for both $d$-wave \cite{Maier:2011, WangF:2011} and $s$-wave \cite{Fang:2011,Mazinb:2011} pairing symmetry.

Experiments have found evidence for the absence of a nodal structure in the superconducting gap.\cite{Wang:2011,Zhang:2011,Li(b):2012,Zeng:2011} This does not identify the gap  symmetry in alkali iron selenides since neither $d$-wave nor $s$-wave symmetry possess any symmetry enforced nodes at $k_z=0$ for this Fermi surface. However, a symmetry based argument has shown that inclusion of $k_z$ dispersion would require the $d$-wave symmetry to possess nodes in the gap structure.\cite{Mazinc:2011} The case for $s$-wave superconducting gap is further supported by angular resolved photoemission (ARPES) measurements in Ref.~\onlinecite{Xu:2012} that find an isotropic superconducting gap structure at the Z point where an electron pocket exists. Scanning tunnelling microscopy (STM) experiments have observed a double gap feature in the local density of states (LDOS).\cite{Li(a):2012,Li(b):2012}

Previous theoretical studies of impurity effects in potassium doped iron selenide superconductors K$_x$Fe$_{2-y}$Se$_2$ have investigated the role of nonmagnetic impurities in identifying the characteristic difference between superconducting gap symmetries.\cite{Zhu:2011,Wang:2013} For example, Zhu {\it et al.}\cite{Zhu:2011} studied the effects of repulsive nonmagnetic impurity potentials using a T-matrix approach for various gap symmetries, and found that sub-gap bound states close to gap edge are generated for $d$-wave but not for $s$-wave (next-nearest neighbor pairing) superconductivity. Another calculation based on a three orbital tight-binding Hamiltonian and the Bogoliubov-de Gennes (BdG) method found that only attractive nonmagnetic impurity potentials generate sub-gap bound states for $d$-wave gap symmetry.\cite{Wang:2013}
This discrepancy between theoretical models for impurity bound states in iron-based superconductors has been recently pointed out to arise naturally for these systems due to strong sensitivity to the particular band structure and superconducting gap function.\cite{Beaird:2012} Therefore, it seems important to use realistic five-band models with self-consistently generated superconducting pairing in order to minimize the effects of free parameters within the various models.

In this work we study the impurity LDOS in the superconducting state of K$_x$Fe$_{2-y}$Se$_2$ using a realistic five-band microscopic model for this material. Superconductivity is introduced into the model by using the effective pairing interactions obtained within the RPA spin-fluctuation exchange mechanism. We focus on nearest neighbor (NN) $d$-wave and next-nearest neighbor (NNN) $s$-wave superconducting states since these are the leading candidates for the ground state gap symmetry in these materials. Note that in the absence of a hole pocket around the $\Gamma$ point, the NNN $s$-wave symmetry does not possess a sign change in the superconducting gap, and the only difference between the two gap symmetries is the absence (presence) of a relative sign difference between the gap on the electron pockets at the M points for $s$-wave ($d$-wave) pairing.

We find that sufficiently strong nonmagnetic scatterers lead to sub-gap bound states both for repulsive and attractive impurity potentials for the case of $d$-wave pairing. The location of the bound states depend rather strongly on the electron doping and approach zero bias for increased electron doping. Hence, we propose to measure the LDOS for maximally electron doped K$_x$Fe$_{2-y}$Se$_2$ samples to clearly identify the possible existence of sub-gap bound states.

For magnetic impurities, bound states exist within the superconducting gap for both $d$- and $s$-wave pairing as expected. It may, however, be too naive to split up the scatterers into nonmagnetic and magnetic since, as we show, in the case of $d$-wave superconductivity putative nonmagnetic impurity potentials may take advantage of significant Coulomb correlations to locally pin magnetic fluctuations and generate magnetic droplets around the impurity sites. This mechanism can be understood as a local Stoner instability and has been discussed extensively in the literature.\cite{tsuchiura01,wang02,zhu02,chen04,kontani06,harter07,andersen07,alloul09,andersen10,christensen11,roemer12,gastiasoro13,gastiasoroprl13} The existence of impurity-induced order may be particularly relevant to alkali doped iron chalcogenides due to the evidence of significant electronic interactions\cite{Dagotto:2012} and the likely presence disordered vacancies (which in the vacancy ordered state induce the BAFM order) in the superconducting regions. We note that the presence of impurity-induced magnetic order is only relevant for $d$-wave pairing, and hence our conclusion of absence (presence) of sub-gap bound states for $d$-wave ($s$-wave) pairing symmetry remains robust.

\begin{figure}[]
\begin{minipage}{.99\columnwidth}
\includegraphics[clip=true,width=0.99\columnwidth]{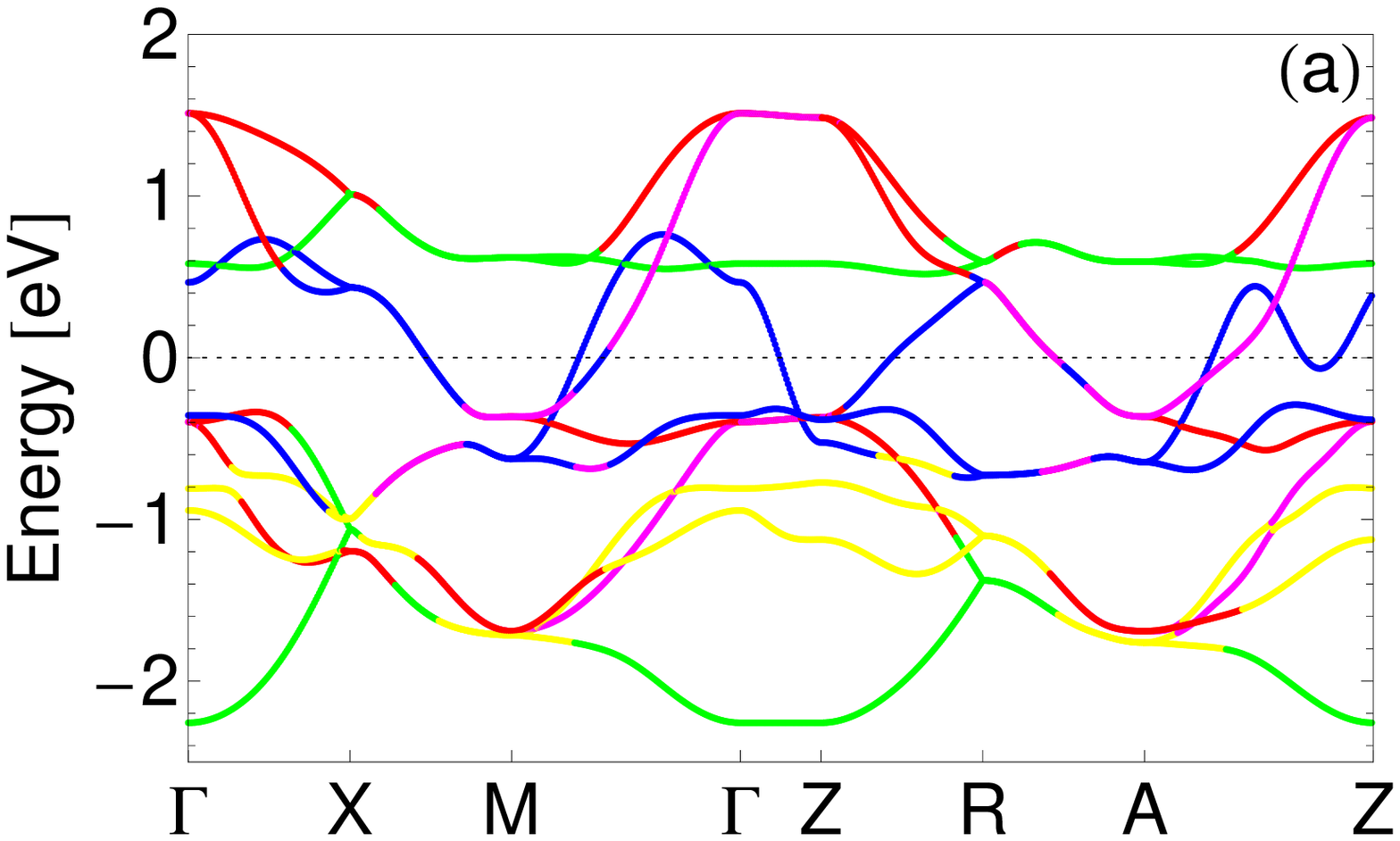}
\end{minipage}
\begin{minipage}{.92\columnwidth}
\includegraphics[clip=true,width=0.92\columnwidth]{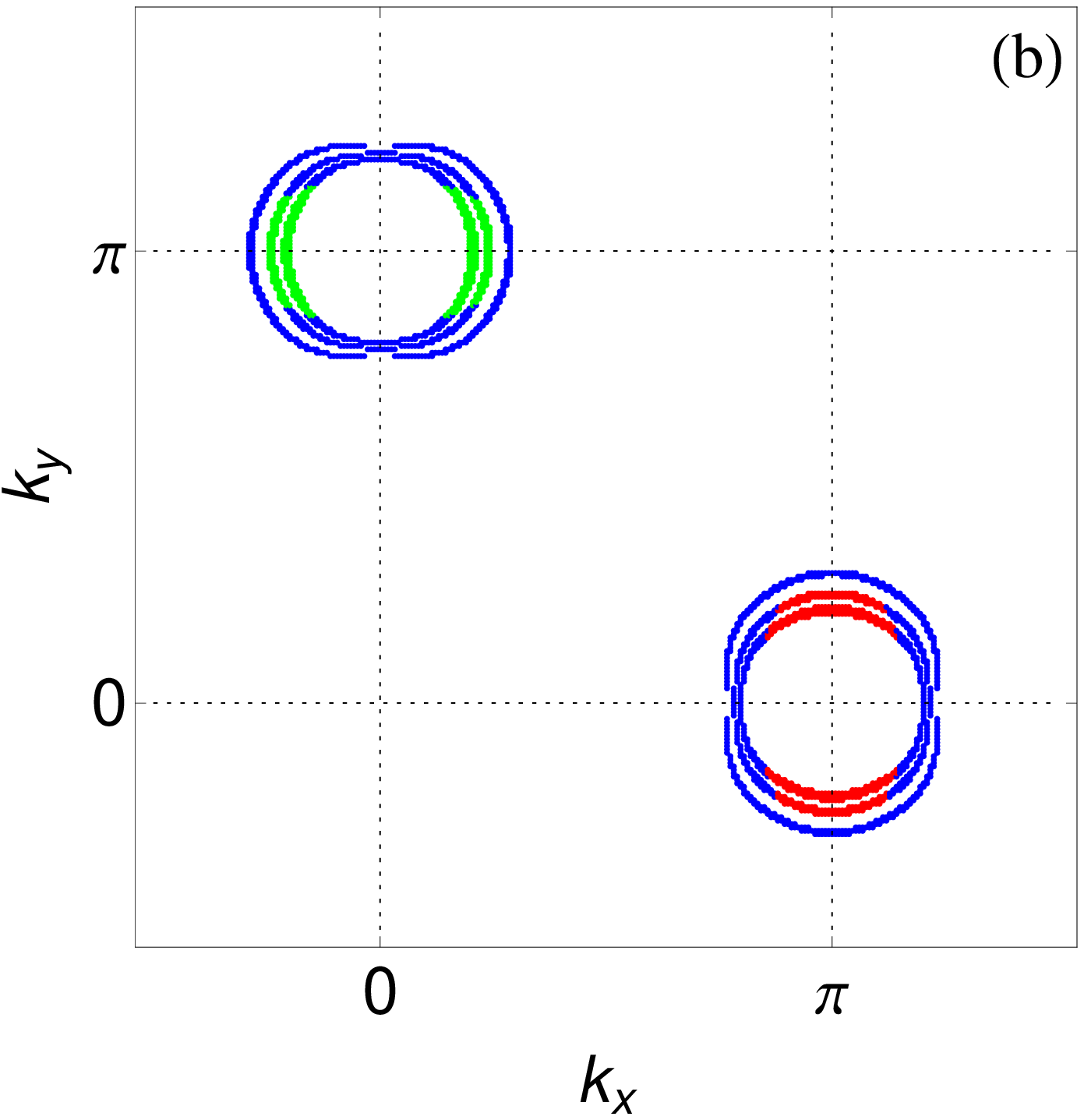}
\end{minipage}
\caption{(Color online) (a) Band structure of K$_x$Fe$_{2-y}$Se$_2$. Here the color codes represent green: $d_{xz}$, red: $d_{yz}$, pink: $d_{x^2-y^2}$, blue: $d_{xy}$, yellow: $d_{z^2}$. (b) Orbitally resolved Fermi surface at $k_z=0$. Outer pockets correspond to doping level $n=6.23$ electrons/Fe, middle pocket to $n=6.19$ electrons/Fe, and inner pocket to $n=6.14$ electrons/Fe.}
\label{fig:normal}
\end{figure}

\section{Model}
For the five orbital model the Nambu space is defined by the creation and annihilation operators $\Psi_{\mu\mathbf{k}}^{\dag}=(c_{\mu\mathbf{k}\uparrow}^{\dag},c_{\mu-\mathbf{k}\downarrow}$),  and $\Psi_{\mu\mathbf{k}}=(c_{\mu\mathbf{k}\uparrow},c_{\mu-\mathbf{k}\downarrow}^{\dag}$), respectively. Here $c_{\mu\mathbf{k}\uparrow}^{\dag} (c_{\mu\mathbf{k}\downarrow})$ denotes the creation (annihilation) operator with orbital index $\mu$ being one of the five $d$-orbitals, $d_{xz}$, $d_{yz}$, $d_{x^2-y^2}$, $d_{xy}$ and $d_{3z^2}$, $\mathbf{k}$ is the momentum index, and $\uparrow (\downarrow)$ is the spin projection. The Hamiltonian for the alkali doped iron chalcogenides in the superconducting state can be expressed in this Nambu basis as
\begin{equation}
\label{eq:H}
\hat{H}(\mathbf{k})=\left(
  \begin{array}{cc}
    \hat{H}_0(\mathbf{k}) & \hat{\Delta}(\mathbf{k}) \\
    \hat{\Delta}^{\dag}(\mathbf{k}) & -\hat{H}^{*}_0(\mathbf{k}) \\
  \end{array}
\right)
\end{equation}
Here $\hat{H}_0(\mathbf{k})$ is a $5\times5$ matrix that represents the tight-binding Hamiltonian
\begin{equation}
 \label{eq:H0}
\hat{H}_{0}(\mathbf{k})=\sum_{\mu\nu,\sigma}t_{\mathbf{k}}^{\mu\nu}c_{\mathbf{k}\mu\sigma}^{\dagger}c_{\mathbf{k}\nu\sigma}-
\mu_{0}\sum_{\mathbf{k}\mu\sigma}n_{\mathbf{k}\mu\sigma},
\end{equation}
where $\mu_{0}$ is the chemical potential and the indices $\mu$ and $\nu$ correspond to the five Fe $d$-orbitals. The hopping parameters $t_{\mathbf{k}}^{\mu\nu}$  include hoppings to the second next-nearest neighbors and are identical to those discussed in Ref.~\onlinecite{Yong:2012,Mukherjee:2013}. To calculate the electronic structure we enforce an electron doping of 0.15 electron per Fe (calculated for the bulk system) by adjusting the chemical potential to $\mu_0=-0.25$eV. The band structure derived from the tight-binding Hamiltonian and the corresponding Fermi surface at $k_z=0$ are shown in Fig.~\ref{fig:normal}(a,b) respectively. The Fermi surface sheets at $k_z=0$ have been plotted in Fig.~(\ref{fig:normal})(b) for three different doping levels corresponding to $n=6.23$ electron/Fe, $n=6.19$ electrons/Fe, and $n=6.14$ electrons/Fe. As seen from Fig.~(\ref{fig:normal})(b), upon increasing the electron doping the shape of the Fermi surface becomes more elliptical and its orbital character gets dominated by the $xy$ orbital.

The off-diagonal block in Eq.(\ref{eq:H}) contains the superconducting part of the Hamiltonian. Here $\hat{\Delta}(\mathbf{k})$ is a $5\times5$ matrix extracted from the quartic interaction term that leads to the superconducting instability
\begin{equation}
 H_{SC}=\sum_{\mu\nu}[\Delta_{\mathbf{k}}^{\mu\nu}
 c_{-\mathbf{k}\mu\uparrow}^{\dagger}c_{\mathbf{k}\nu\downarrow}^{\dagger}+H.c.],
\end{equation}
with the superconducting gap function $\Delta_{\mathbf{k}}^{\mu\nu}$ defined as
\begin{equation}
\Delta_{\mathbf{k}}^{\mu\nu}=\sum_{\mathbf{k'}}\Gamma_{\mu,\nu}^{\mu,\nu}(\mathbf{k},\mathbf{k'})\langle c_{\mathbf{k'}\nu\downarrow}c_{\mathbf{-k'}\mu\uparrow}\rangle.
\end{equation}

\begin{figure}[]
\begin{minipage}{.99\columnwidth}
\includegraphics[clip=true,width=0.99\columnwidth]{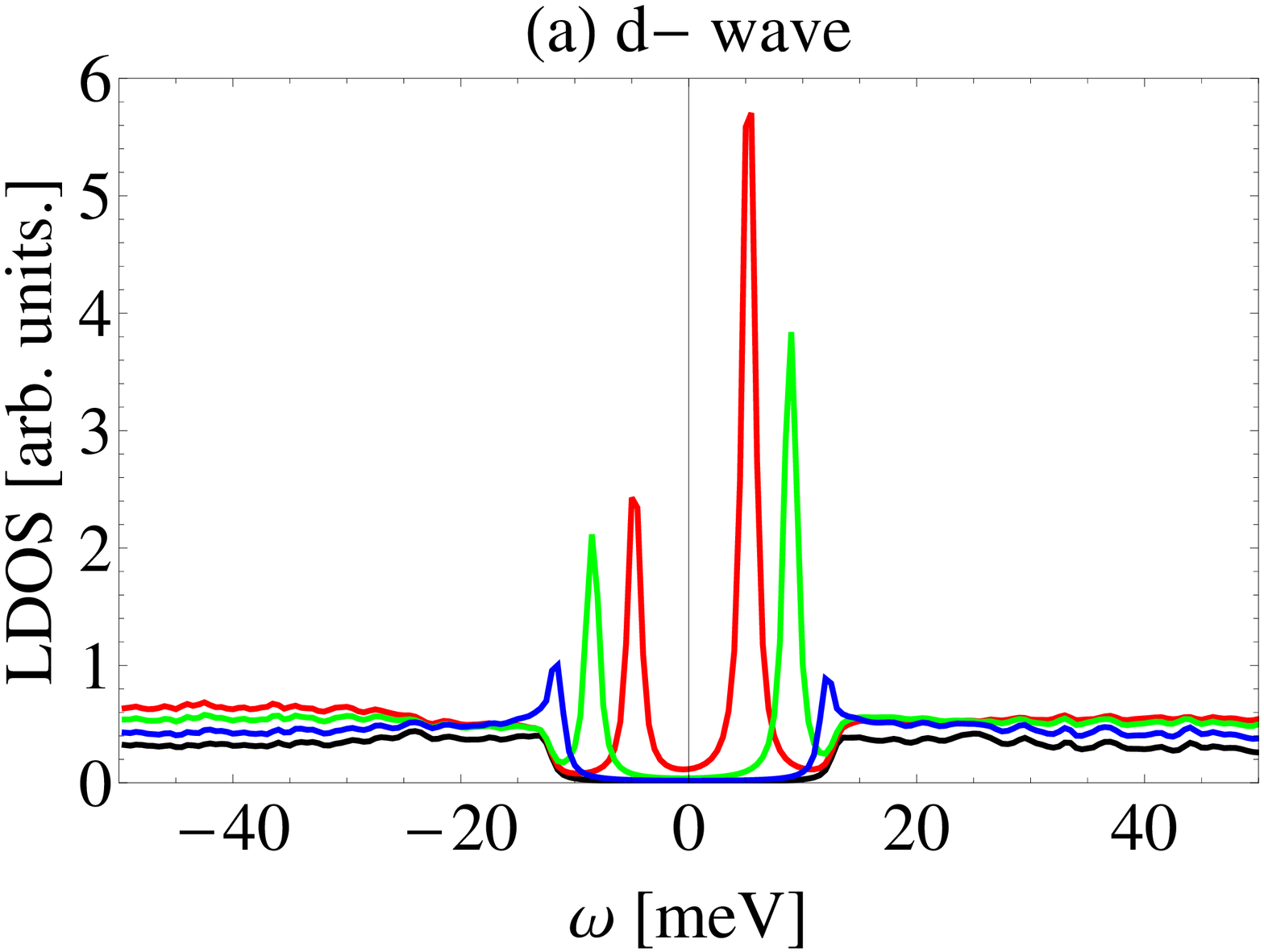}
\end{minipage}
\begin{minipage}{0.99\columnwidth}
\includegraphics[clip=true,width=1.07\columnwidth]{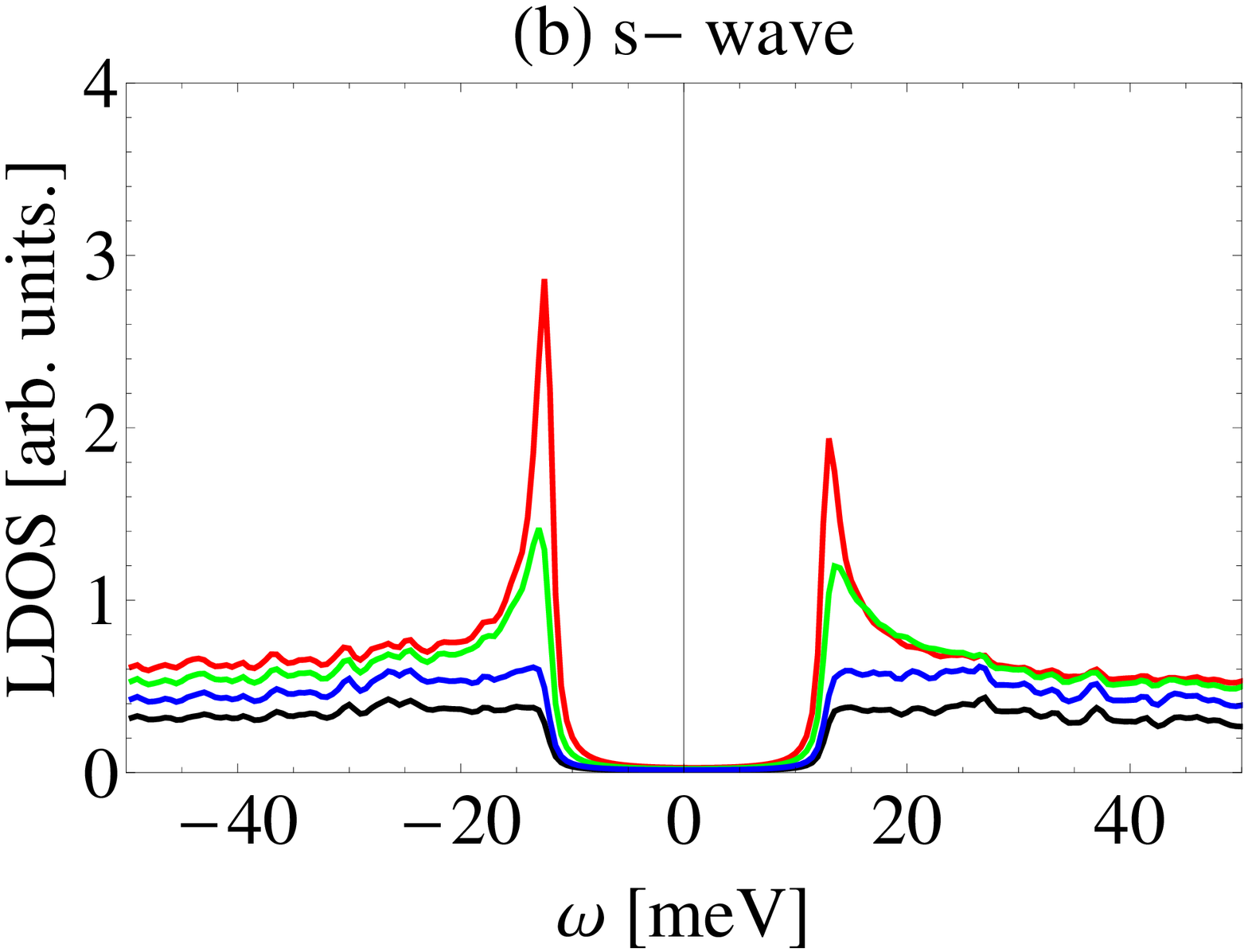}
\end{minipage}
\caption{(Color online) LDOS for nearest neighbor sites around a nonmagnetic impurity for repulsive potentials $V_{imp}$ at $n=6.19$ electrons/Fe in the case of $d$-wave (a) and $s$-wave (b) pairing. Here the color codes represent red: $V_{imp}=10^3$eV, green: $V_{imp}=10$eV, blue: $V_{imp}= 3$eV, black: homogeneous case.} 
\label{fig:nonmagpos}
\end{figure}


Here, $\Gamma_{\mu,\nu}^{\mu,\nu}(\mathbf{k},\mathbf{k'})$ is the pairing interaction obtained from the spin-fluctuation exchange mechanism as mentioned above. The details of the calculations of the pairing interaction and superconducting gap value can be found in Ref.~\onlinecite{Mukherjee:2013}. As discussed in Ref.~\onlinecite{Mukherjee:2013} the LDOS for the homogeneous superconducting state evaluated from this model reveals a nodeless gap for both $d$-wave and $s$-wave superconducting pairing symmetry.  

We solve the single impurity problem in K$_x$Fe$_{2-y}$Se$_2$ superconductors using the so-called T-matrix approach. This requires evaluation of the bare Greens function in Nambu space using the Hamiltonian in Eq.(\ref{eq:H})

\begin{equation}
\label{eq:baregreen}
\hat{G}^{-1}_0(\mathbf{k},i\omega_n)=i\omega_n\hat{I}-\hat{H}(\mathbf{k}),
\end{equation}
where $\omega_n=(2n+1)\pi T$ is the Matsubara frequency for fermions. In the following we assume that the impurities substitute iron atoms
and act as short-ranged potential scatterers.\cite{andersen06} The full Greens function in real-space is given by
\begin{equation}
\label{eq:fullgreen}
\hat{G}(i,j;i\omega_n)=\hat{G}_0(i-j;i\omega_n)+\hat{G}_0(i;i\omega_n)\hat{T}(i\omega_n)\hat{G}_0(-j;i\omega_n).
\end{equation}
Here, the Greens function is evaluated by Fourier transformation $\hat{G}_0(i,j;i\omega_n)={1\over N_xN_y}\sum_{\mathbf{k}}\hat{G}_0(\mathbf{k};i\omega_n)e^{i\mathbf{k}\cdot(\mathbf{r}_i-\mathbf{r}_j)}$ with $N_xN_y$ denoting the total number of lattice sites.
The $\hat{T}$-matrix is given by
\begin{equation}
\label{eq:tmatrix}
\hat{T}(i\omega_n)=\hat{H}_{imp}[\hat{I}-\hat{G}_0(0,0;i\omega_n)\hat{H}_{imp}]^{-1},
\end{equation}
where the impurity potential Hamiltonian is given by $\hat{H}_{imp}=V_{imp}\Psi_0^{\dag}\tau_3\Psi_0$ for nonmagnetic scatterers and $\hat{H}_{imp}=V_{mag}\Psi_0^{\dag}\tau_0\Psi_0$ for magnetic scatterers. Here, $\tau=(\tau_1,\tau_2,\tau_3)$ denote Pauli spin matrices in the $10\times10$ Nambu space and $\tau_0$ is the identity matrix. For simplicity we include only the effects of constant intraorbital impurity potentials. 

It can be instructive to compare the results obtained numerically from the multi-band T-matrix approach with the corresponding single-band analytic expressions to identify the role of multi-band physics in these systems. The bound state energy can be obtained from the T-matrix given in Eq.(\ref{eq:tmatrix}) by evaluating the zeroes of $\det[\hat{I}-\hat{G}_0(0,0;i\omega_n)\hat{H}_{imp}]$. For a single-band problem the bare Greens function is a $2\times2$ matrix with components
\begin{eqnarray}
\label{eq:singreen}
\Re[G_{0,11/22}(0,0;\omega)]&=&\sum_{\mathbf{k}}{\omega \pm \xi(\mathbf{k})\over \omega^2-\xi(\mathbf{k})^2-\Delta^2} \nonumber\\                                                &=&\gamma(\omega) \omega \mp \alpha\\
\label{eq:singreen2}
\Re[G_{0,12/21}(0,0;\omega)]&=&\sum_{\mathbf{k}}{\Delta \over \omega^2-\xi^2(\mathbf{k})-\Delta^2}.
\end{eqnarray}
Here, $\xi(\mathbf{k})$ is the single-particle dispersion and $\Delta$ the superconducting gap.
We have introduced $\gamma(\omega)=-{\pi N(0) \omega \over \sqrt{\Delta^2-\omega^2}} $ with $N(0)$ denoting the density of states at the Fermi level. The parameter $\alpha$ depends on details of the band structure and anisotropy of the superconducting gap over the Fermi surface. For the simplest case of an isotropic superconducting gap, $\alpha$ is given by the expression
\begin{eqnarray}
\label{eq:alpha}
\alpha={N(0)\over 2} \log\left({B^2+\Delta^2-\omega^2\over A^2 +\Delta^2 -\omega^2}\right)
\end{eqnarray}
Where $B$ and $A$ denote the maximum and minimum of the band with respect to the chemical potential.

The off-diagonal parts of the bare Greens function vanish  $\Re[G_{0,12/21}(0,0;\omega)]=0$ for a $d$-wave superconducting gap since $\sum_{\mathbf{k}} \Delta(\mathbf{k})=0 $. This occurs because although the gap over a Fermi pocket does not change sign for $d$-wave symmetry it changes sign between the two electron pockets at the M points. For an $s$-wave symmetry the off-diagonal term is given by $\Re[G_{0,12/21}(0,0;\omega)]=\gamma(\omega) \Delta$. Using the above formulation the bound state energy $\omega_B$ for a nonmagnetic impurity is given by,

\begin{equation}
\label{eq:nonmagcalc}
\omega_B^2 = {\omega_0^2+p\over \omega_0^2+1}\Delta^2,
\end{equation}
with $p=0$ for $d$-wave and $p=1$ for $s$-wave symmetry. The impurity potential is included through the dimensionless term $\omega_0^2={1\over \pi^2N(0)^2}(1/V_{imp}+\alpha)^2$.

In the case of a magnetic impurity potential, the singularity condition for the T-matrix leads to the following expression for bound state energy given by,
\begin{equation}
\label{eq:magcalc}
\omega_B^2={-(1-\lambda)(\lambda-p)-x \pm \sqrt{x^2-2x(\lambda-p)(1-p)}\over (1-\lambda)^2+2x}\Delta^2,
\end{equation}
where $x= {2\over \pi^2N(0)^2V^2_{mag}}$, and $\lambda={1\over \pi^2N(0)^2}(1/V_{mag}^2-\alpha^2)$.

In the next section we discuss the results of LDOS around nonmagnetic and magnetic impurities using the numerical five band T-matrix approach and compare the results with the analytic expressions derived in Eq.~(\ref{eq:nonmagcalc}) and Eq.~(\ref{eq:magcalc}).

\begin{figure}[]
\begin{minipage}{.99\columnwidth}
\includegraphics[clip=true,width=0.99\columnwidth]{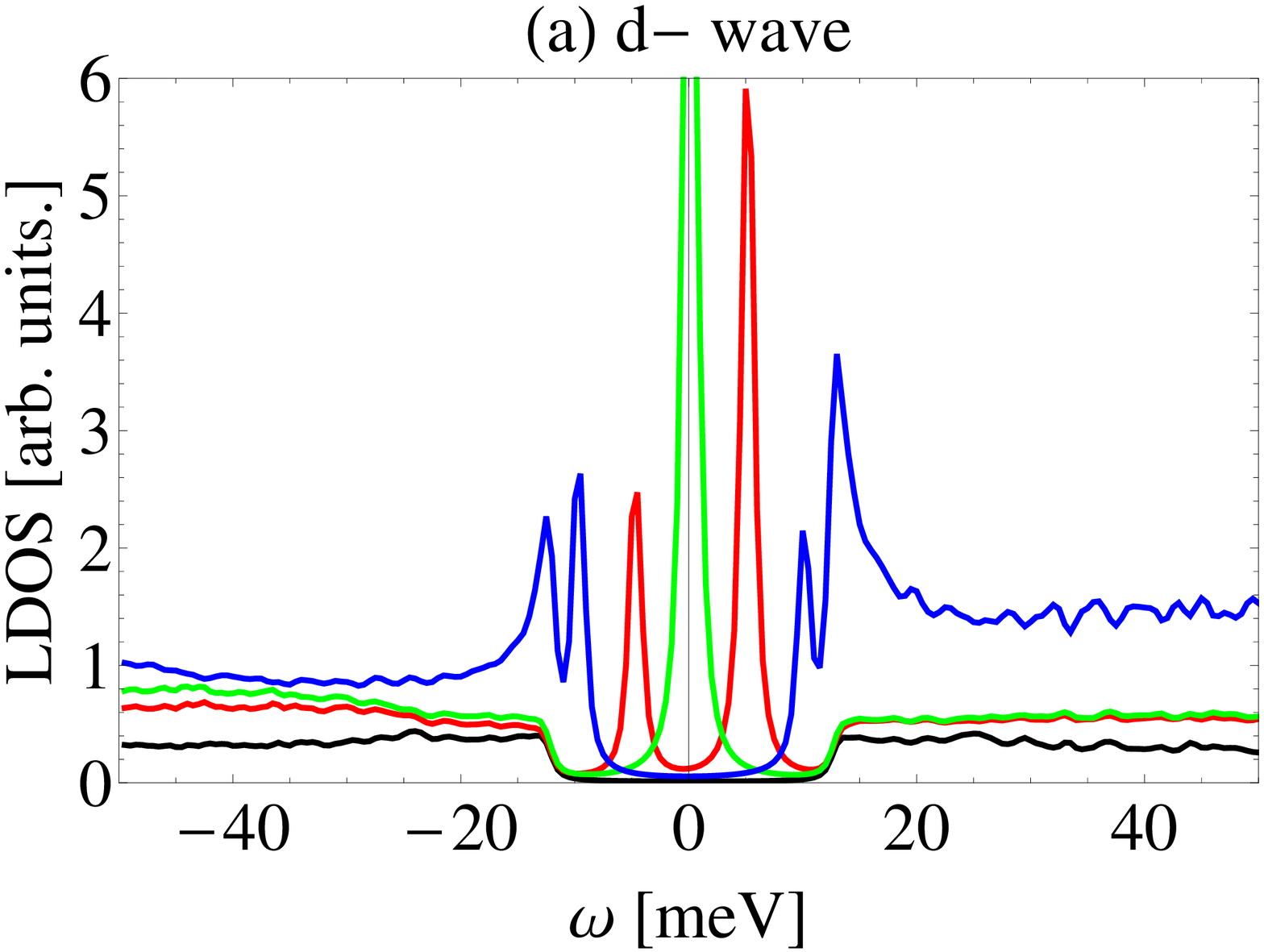}
\end{minipage}
\begin{minipage}{.99\columnwidth}
\includegraphics[clip=true,width=1.07\columnwidth]{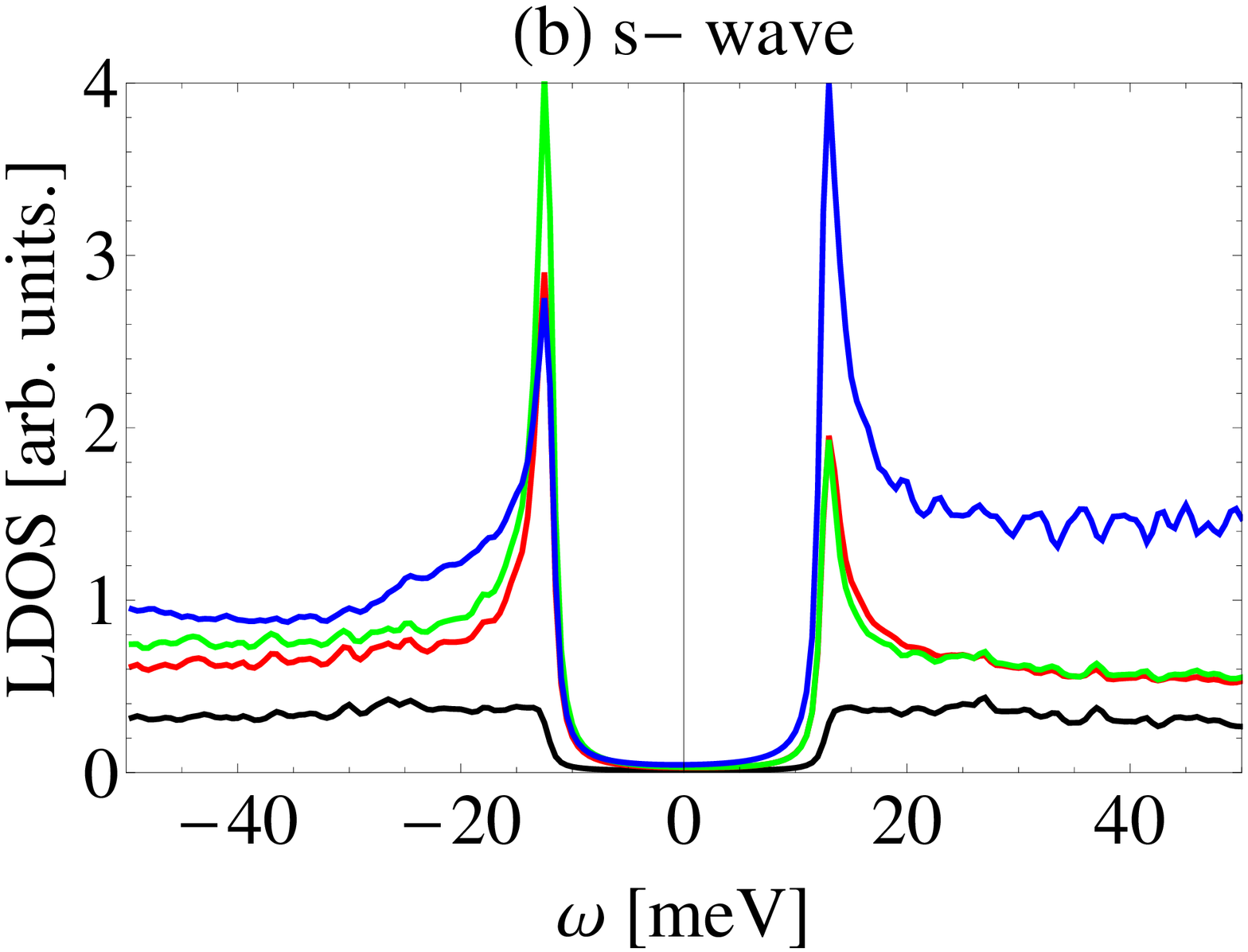}
\end{minipage}
\caption{(Color online) LDOS for nearest neighbor sites around a nonmagnetic impurity for attractive impurity potentials $V_{imp}$ at $n=6.19$ electrons/Fe in the case of $d$-wave (a) and $s$-wave (b) pairing. The color codes represent red: $V_{imp}=-10^3$eV, green: $V_{imp}=-10$eV, blue: $V_{imp}= -3$eV, black: homogeneous case.}\label{fig:nonmagneg}
\end{figure}

\section{Results}

The LDOS for the nearest neighbor sites around nonmagnetic impurities are shown in Fig.~\ref{fig:nonmagpos} and Fig.~\ref{fig:nonmagneg} for repulsive and attractive potentials, respectively. For repulsive impurity potentials we find that sub-gap bound states exist only in the presence of a $d$-wave superconducting gap in general agreement with Anderson's theorem. For systems with an electron doping of $n=6.19$ electrons/Fe at $k_z=0$ (which corresponds to $n=6.15$ electrons/Fe for the bulk ($k_z$-summed) system) the LDOS around large repulsive potentials such as iron vacancies shows sub-gap bound state peaks around $\omega_B/\Delta \sim 0.55$. The bound state peak position saturates around this value of $\omega_B/\Delta \sim 0.55$ and does not shift further into the superconducting gap for larger impurity potentials. It can also be seen from Fig.~\ref{fig:nonmagpos}(a) that the bound state peak positions move towards the gap edge for weaker repulsive potentials in agreement with earlier calculations by Zhu {\it et al.}\cite{Zhu:2011}
\begin{figure}[]
\begin{minipage}{.99\columnwidth}
\includegraphics[clip=true,width=0.99\columnwidth]{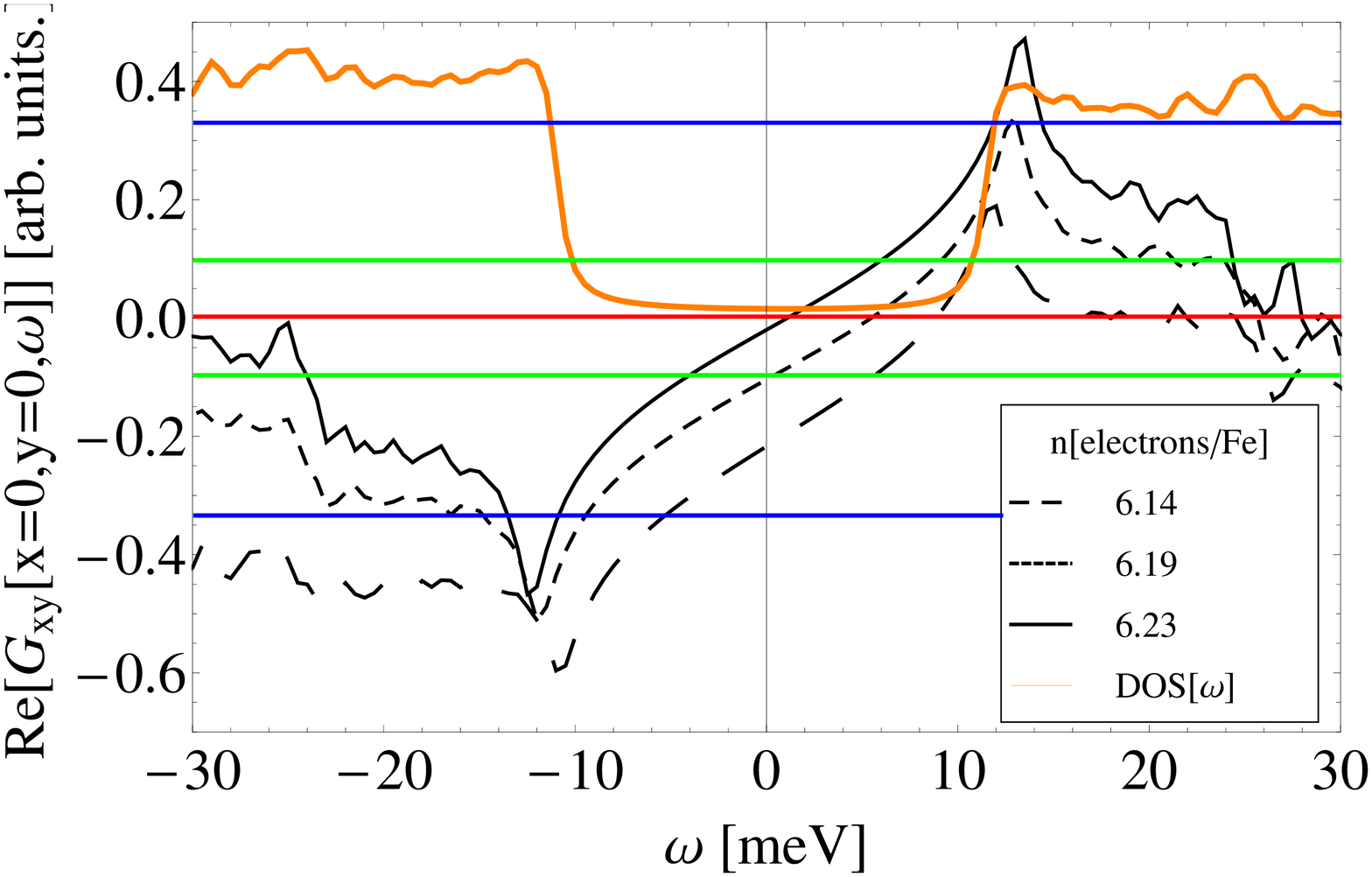}
\end{minipage}
\caption{(Color online) $Re[G_{0,xy}(x=0,y=0,\omega)]$ and DOS versus $\omega$ in the superconducting state for different electron doping levels. The sub-gap bound state positions for nonmagnetic impurities are well represented by the condition $Re[G_{0,xy}(x=0,y=0,\omega)]=1/V_{imp}$. Horizontal lines represent $1/V_{imp}$ values with colors red: $V_{imp}=\pm10^3$eV, green: $V_{imp}=\pm10$eV, blue: $V_{imp}= \pm3$eV.} \label{fig:realgreen}
\end{figure}

For attractive impurity potentials the sub-gap bound states are also found to exist only for $d$-wave gap symmetry. As can be seen in Fig.~\ref{fig:nonmagneg}(a), the bound state peak positions in this case move towards zero bias for impurity potential of $V_{imp} \sim -10$eV. Note that the bound state peak position saturates at $\omega_B/\Delta \sim 0.55$ for large impurity potentials even when the potential is attractive.

The location of the resonant peaks in the $d$-wave superconducting state for the entire range of nonmagnetic impurity potentials can be understood from Fig.~\ref{fig:realgreen} which displays $Re[G_{0,xy}(x=0,y=0,\omega)]$ as a function of $\omega$ for the same three doping levels as in Fig.~\ref{fig:normal}(b). Note that $xy$ index in the bare greens function refers to its diagonal component corresponding to the $xy$ orbital. As seen from the cuts of the green, blue and red impurity potential line on the short dashed curve representing electron density of 6.19 electrons/Fe, the resonant peaks in Fig.~\ref{fig:nonmagpos} lie at positions corresponding to $\omega$ values where $1/V_{imp}=Re[G_{0,xy}(x=0,y=0,\omega)]$. The good correspondence between the sub-gap peak position and the $\omega$ value given by $1/V_{imp}=Re[G_{0,xy}(x=0,y=0,\omega)]$ for a $d$-wave superconducting gap is due to the condition $\sum_k \Delta_k=0$ which makes the bare Greens function matrix diagonal.

Our results for attractive impurity potentials agree qualitatively with those by Wang {\it et al.}\cite{Wang:2013} but unlike their results we also find sub-gap bound states for repulsive potentials for $d$-wave superconductivity as evident from Fig.~\ref{fig:nonmagpos}(a). Such discrepancies could arise from variations in band structure or due to differences in chemical doping. In Fig.~\ref{fig:realgreen} we also show $Re[G_{0,xy}(x=0,y=0,\omega)]$ for doping levels $n=6.14$ and $n=6.23$ electrons/Fe, respectively. As can be seen by intersection of the bare Greens function with $1/V_{imp}$, for $n=6.23$ the bound state peaks for a given repulsive impurity potential move closer to zero bias compared to $n=6.19$. However, for lower electron doping of $n=6.14$ the sub-gap bound states are pushed to the gap edge for all repulsive impurity potentials even for a $d$-wave superconducting state, and can exist within the $d$-wave superconducting gap only for attractive impurity potentials. The significant shift of the bound state peak position with chemical doping results from a change in electron concentration as well as a change in shape and orbital content of the Fermi surface as shown in Fig.\ref{fig:normal}(b).

\begin{figure}[]
\begin{minipage}{.99\columnwidth}
\includegraphics[clip=true,width=0.99\columnwidth]{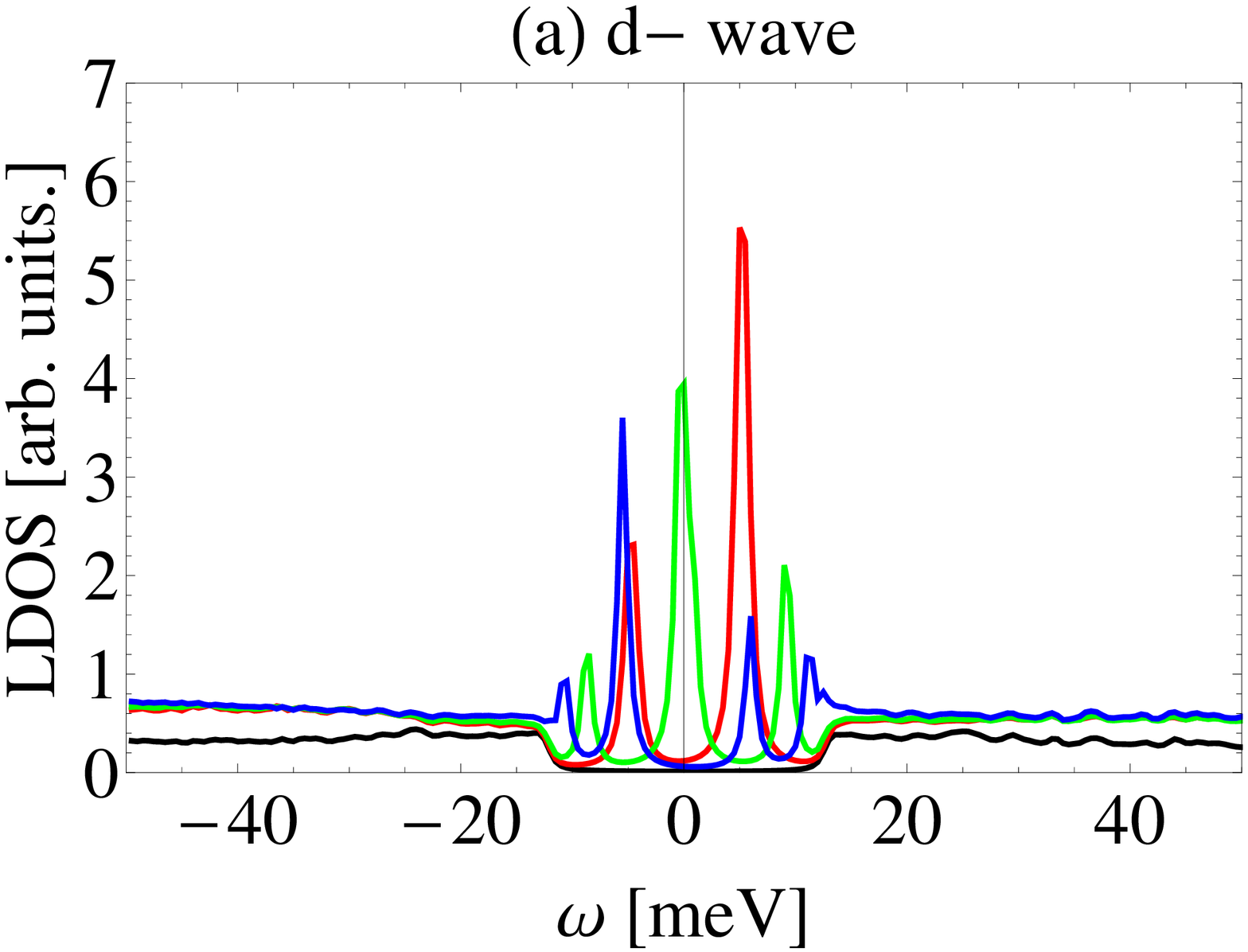}
\end{minipage}
\begin{minipage}{.99\columnwidth}
\includegraphics[clip=true,width=0.99\columnwidth]{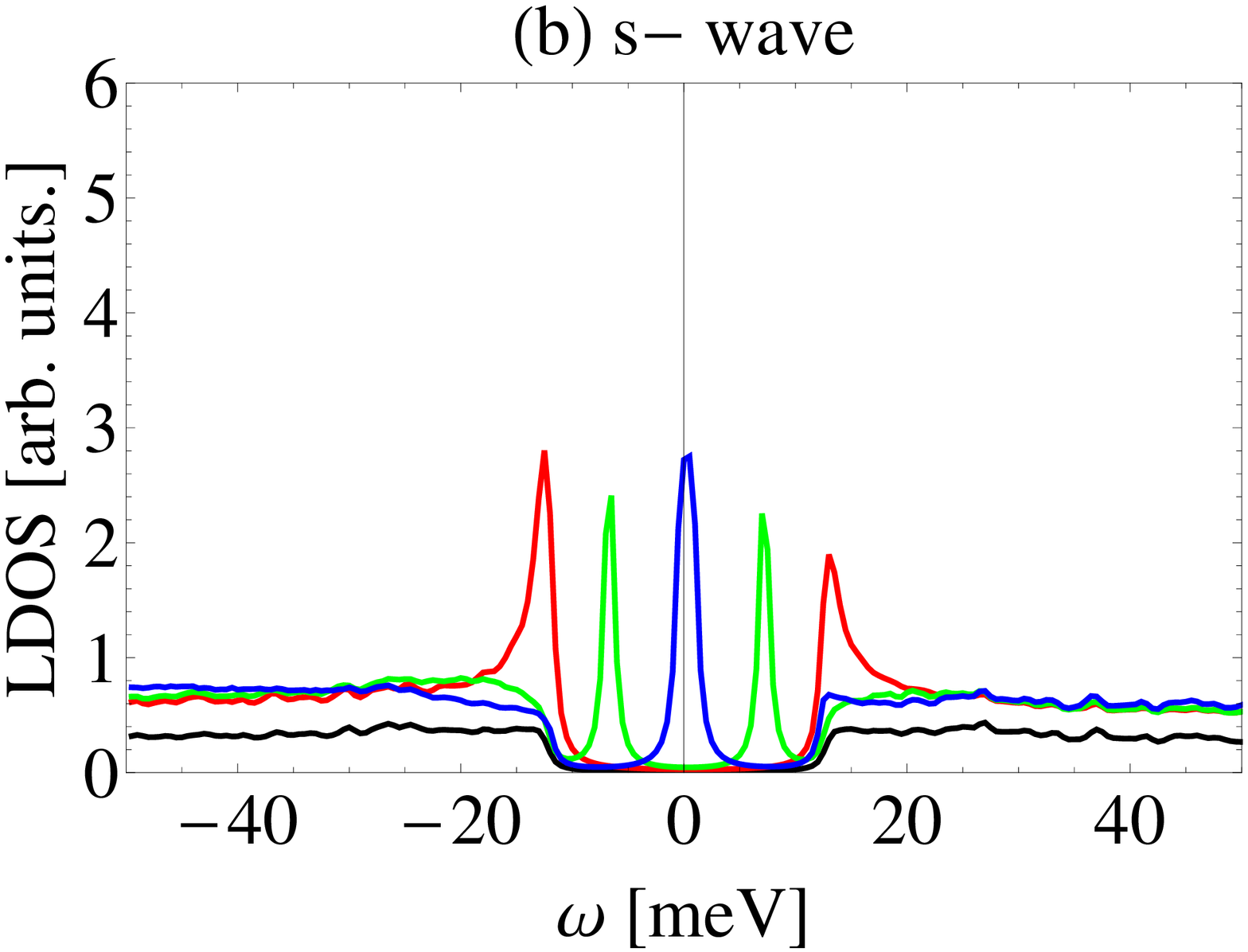}
\end{minipage}
\caption{(Color online) LDOS for nearest neighbor sites for an intraorbital magnetic impurity potential $V_{mag}$ at $n=6.19$ electrons/Fe. The color codes represent red: $V_{mag}=10^3$eV, green: $V_{mag}=10$eV, blue: $V_{mag}= 5$eV, and black: homogeneous DOS.} \label{fig:magimp}
\end{figure}

\begin{figure}[]
\begin{minipage}{.99\columnwidth}
\includegraphics[clip=true,width=0.99\columnwidth]{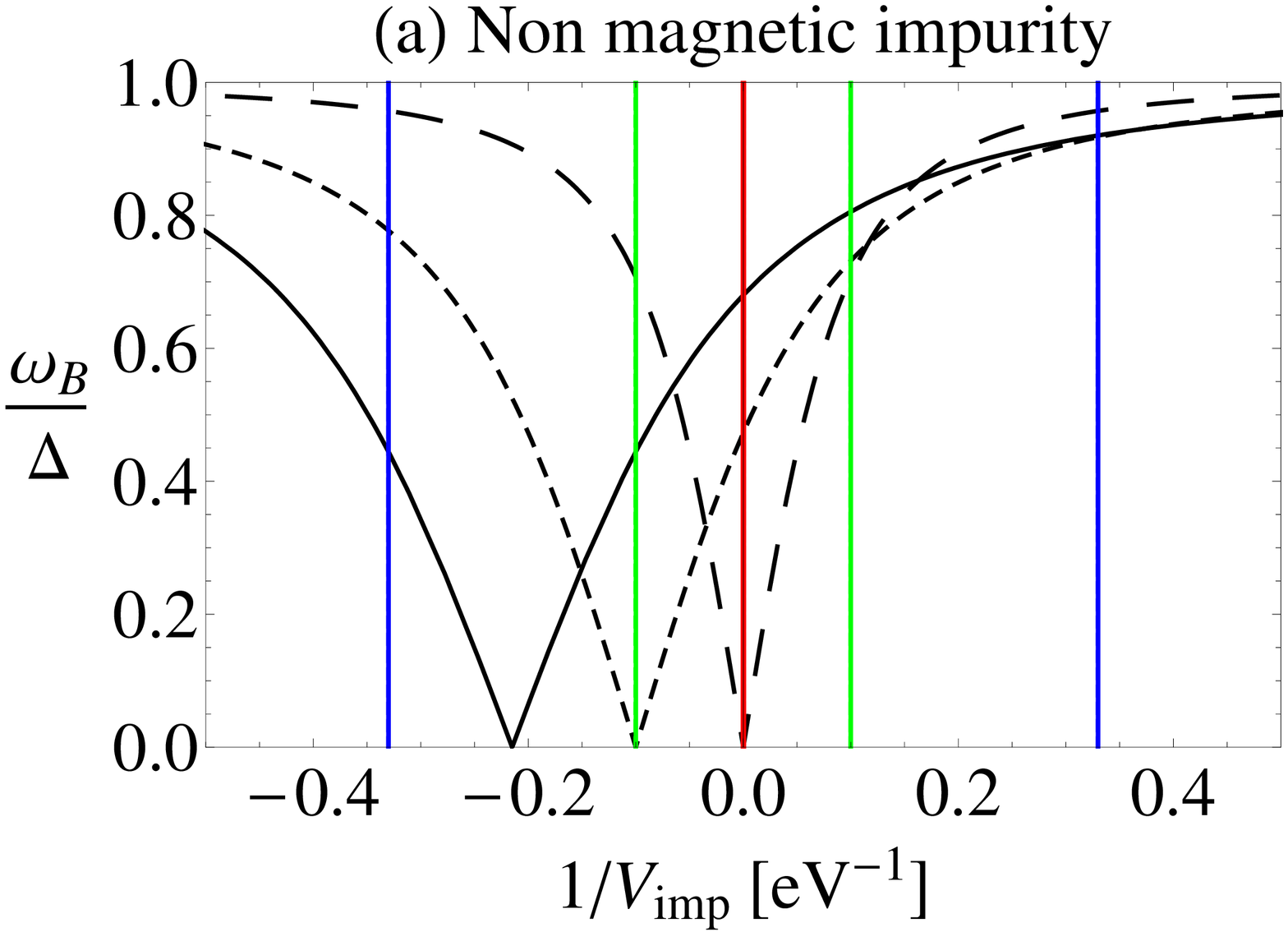}
\end{minipage}
\begin{minipage}{.49\columnwidth}
\includegraphics[clip=true,width=0.99\columnwidth]{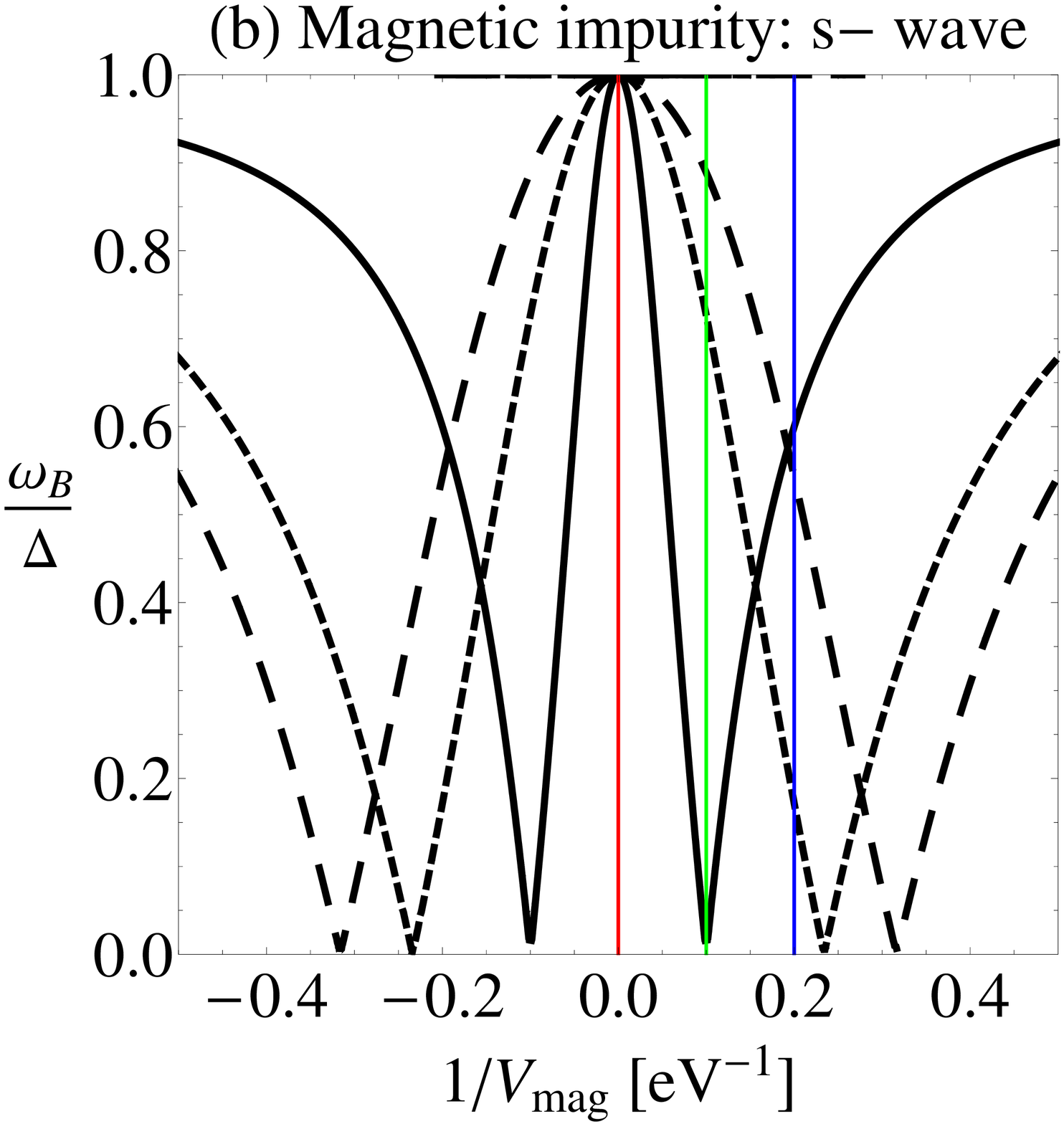}
\end{minipage}
\begin{minipage}{.49\columnwidth}
\includegraphics[clip=true,width=0.99\columnwidth]{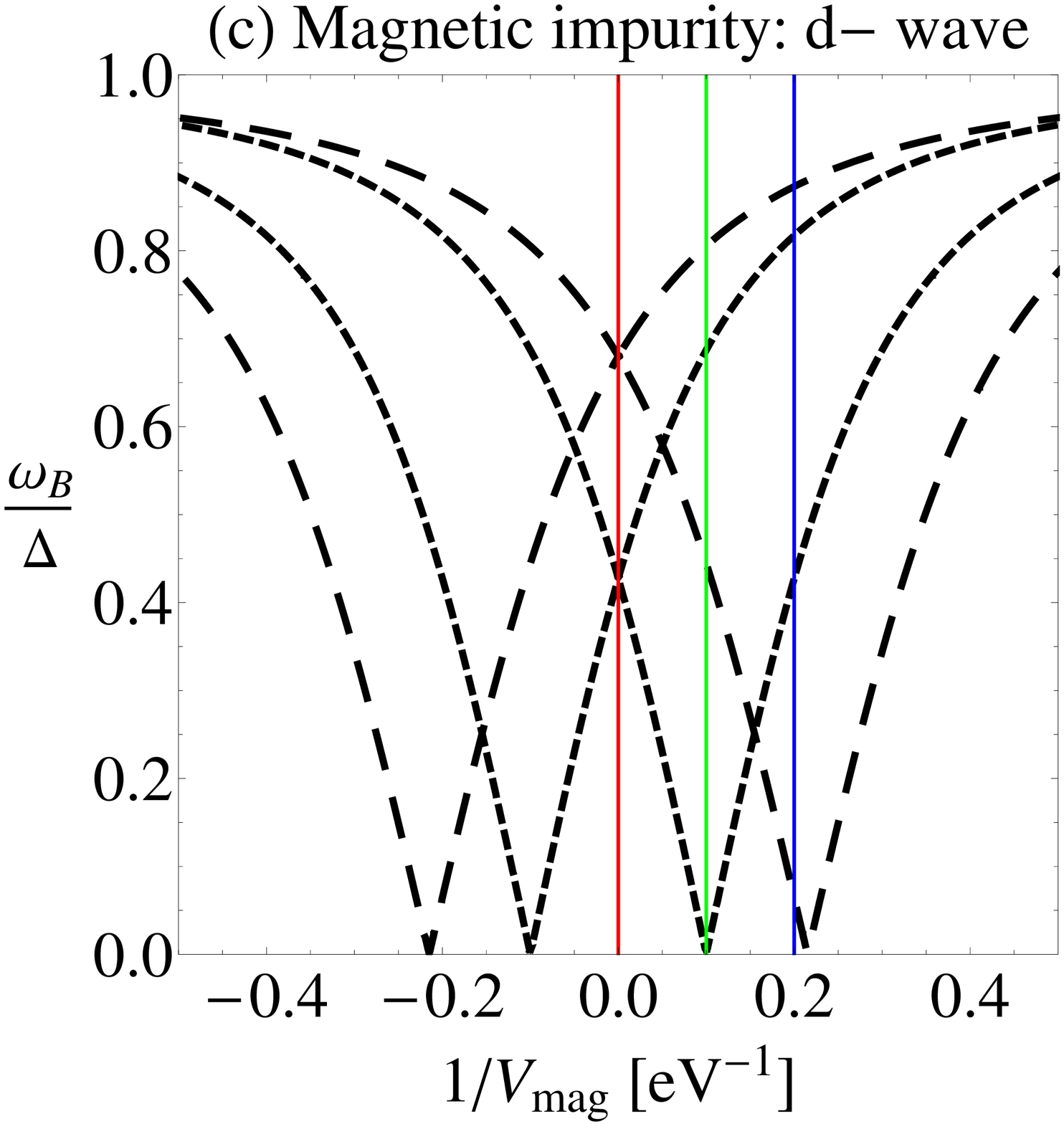}
\end{minipage}
\caption{(Color online) Position of the sub-gap bound state energy versus $1/V_{imp}$ obtained using Eq.~(\ref{eq:nonmagcalc}) and Eq.~(\ref{eq:magcalc}) for different electron doping levels. (a) Nonmagnetic impurities in the case of $d$-wave pairing symmetry. (b) Magnetic impurity in $s$-wave superconductor. (c) Magnetic impurity in $d$-wave superconductor. No sub-gap bound states exist for doping level of $n=6.23$ electrons/ Fe. Plot symbols for the different doping level are same as those used in Fig.\ref{fig:realgreen}. Vertical lines show position of impurity potential with colors corresponding to V$_{imp}$ values used in Fig.\ref{fig:nonmagpos}, Fig.\ref{fig:nonmagneg}, and Fig.\ref{fig:magimp}. }
 \label{fig:impcalc}
\end{figure}

In order to better understand the significant shifts in the bound state peak positions with changes in chemical doping we can compare the numerical results in Fig.~\ref{fig:nonmagpos}, and Fig.~\ref{fig:nonmagneg} with the analytic expression given in Eq.~(\ref{eq:nonmagcalc}). We find that the single-band analytic form for the parameter $\alpha$ given in Eq.~(\ref{eq:alpha}) cannot describe the strong shift in bound state peak positions with changes in impurity potential or chemical doping for nonmagnetic impurities. However since the Fermi surface and the normal state DOS are dominated by the $d_{xy}$ orbital, we introduce an effective single-band model with renormalized parameters $\alpha$ and $\pi N(0)$. Specifically we take $\alpha$ and $\pi N(0)$ as free parameters and fix their values such that the bound state energy obtained from the analytic expression in Eq.~(\ref{eq:nonmagcalc}) agrees with the full 5 orbital T-matrix calculation for $V_{imp}=10^3$eV and $V_{imp}=10$eV. Applying this procedure for different electron densities we get $1/(\pi N(0))= 4.728$eV and $1/\alpha=10$eV for $n=6.19$ electrons/ Fe, $n=6.23$ corresponds to $1/(\pi N(0))= 10$eV, $\alpha=0$, and $n=6.14$ gives $1/(\pi N(0))= 4.313$eV, $1/\alpha=4.651$eV. Having obtained the parameter values for $\alpha$ and $\pi N(0)$ we now test the correspondence between the analytic expressions in Eq.~(\ref{eq:nonmagcalc}) and Eq.~(\ref{eq:magcalc}) with the numerical solution for arbitrary impurity potentials. In Fig.~\ref{fig:impcalc}(a) the plot of bound state peak position $\omega_B$ obtained from Eq.~(\ref{eq:nonmagcalc}) is shown against $1/V_{imp}$ for different electron densities in the presence of nonmagnetic impurities. Interestingly, the bound state positions evaluated from the analytic expression in Eq.~(\ref{eq:nonmagcalc}) agree quantitatively with the results obtained from the numerical calculation of the full five band T-matrix approach for all values of nonmagnetic impurity potentials including the results presented in Fig.~\ref{fig:nonmagpos} and Fig.~\ref{fig:nonmagneg}. We attribute this agreement with the functional form of a single-band model to the dominant role of the $d_{xy}$ orbital in the normal state DOS and the orbital resolved superconducting gap magnitudes.\cite{Mukherjee:2013}

LDOS calculations have also been performed in the presence of local magnetic impurities. The results are shown in Fig.~\ref{fig:magimp}(a,b). It is found that impurity induced multiple sub-gap bound states exist for both $d$-wave and $s$-wave superconducting gap symmetries again in overall agreement with Anderson's theorem. These results hold significance for the alkali iron chalcogenides since the superconducting state exists in phase separated regions from a strong BAFM state. Multiple bound state peaks have been found in the LDOS calculation near an interface between a superconductor and a $\sqrt{5}\times \sqrt{5}$ vacancy ordered BAFM state in K$_x$Fe$_{2-y}$Se$_2$ for both $s$-wave and $d$-wave superconducting states in Ref.~\onlinecite{Mukherjee:2013}. Though the vacancies act as large nonmagnetic impurity potentials and are expected to generate sub-gap bound state peaks only for a $d$-wave pairing as discussed above, the presence of BAFM magnetism leads to sub-gap bound states for both $d$-wave and $s$-wave superconducting state. It is this collective influence of vacancy potential and magnetism that explains the sub-gap bound state peaks for both $d$-wave and $s$-wave superconducting states found near interfaces between superconducting and BAFM regions in Ref.~\onlinecite{Mukherjee:2013}

The numerically obtained bound state peak positions can be again compared to the analytic expression for a single-band model. The solutions of Eq.~(\ref{eq:magcalc}) are plotted in Fig.~\ref{fig:impcalc}(b,c) for the same parameters $\pi N(0)$ and $\alpha$ that were obtained for the case of nonmagnetic impurities. The analytic solution agrees with the numerical results for all magnetic impurity potentials including the results presented in Fig.~\ref{fig:magimp}. The sub-gap peak position plots in Fig.~\ref{fig:impcalc}(b,c) also reveal that magnetic impurities can lead to multiple bound states. For an $s$-wave state shown in Fig.~\ref{fig:impcalc}(b) we never find more than two sub-gap bound states whereas $d$-wave states generally form four sub-gap peaks as seen directly from Fig.~\ref{fig:impcalc}(c). However this difference is observable only for strong magnetic impurity potentials since the multiple bound state peaks are pushed towards the superconducting gap edge and become difficult to resolve for weaker magnetic impurity potentials.

 \begin{figure}[]
\begin{minipage}{.99\columnwidth}
\includegraphics[clip=true,width=0.99\columnwidth]{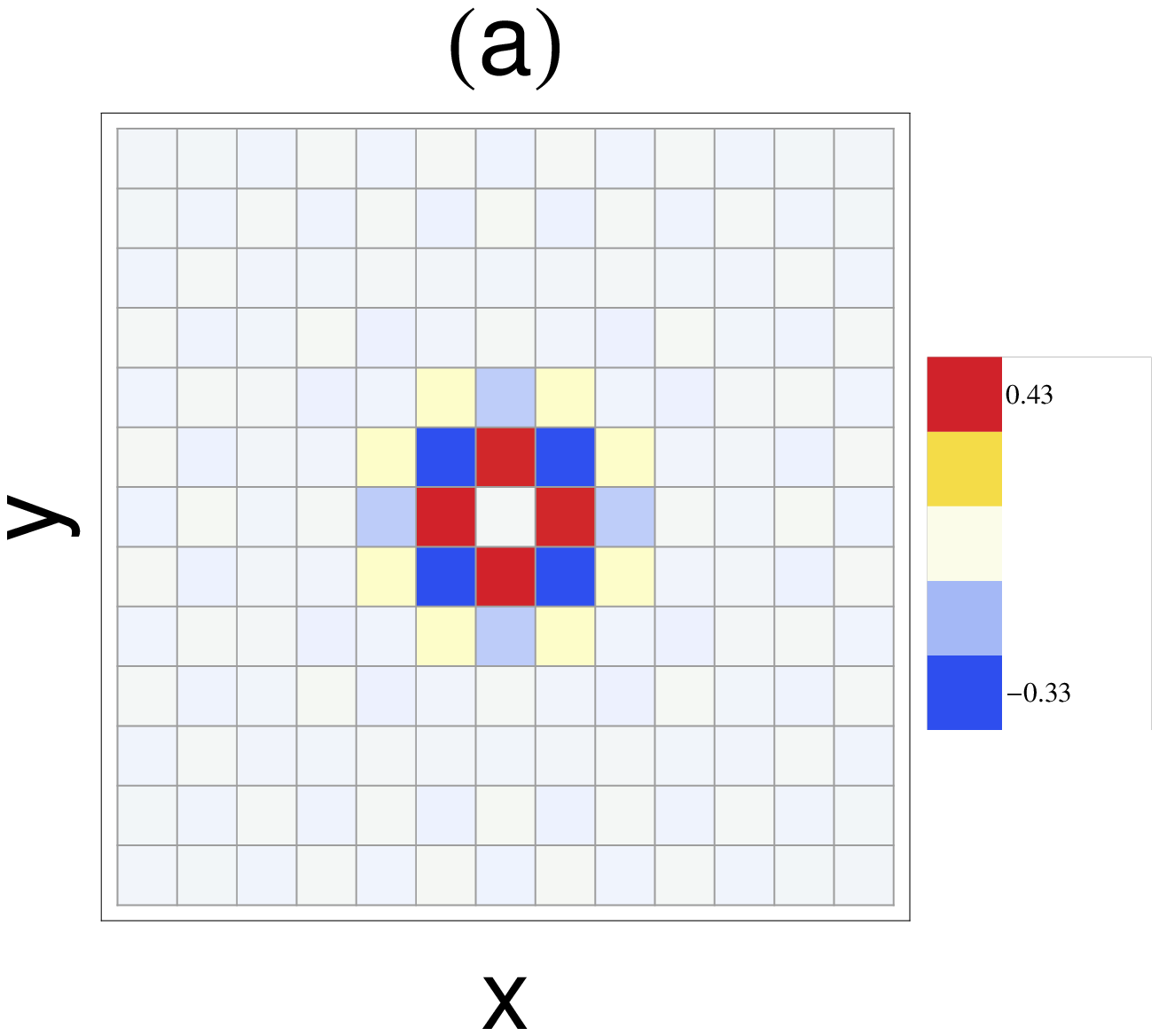}
\end{minipage}
\begin{minipage}{.99\columnwidth}
\includegraphics[clip=true,width=0.99\columnwidth]{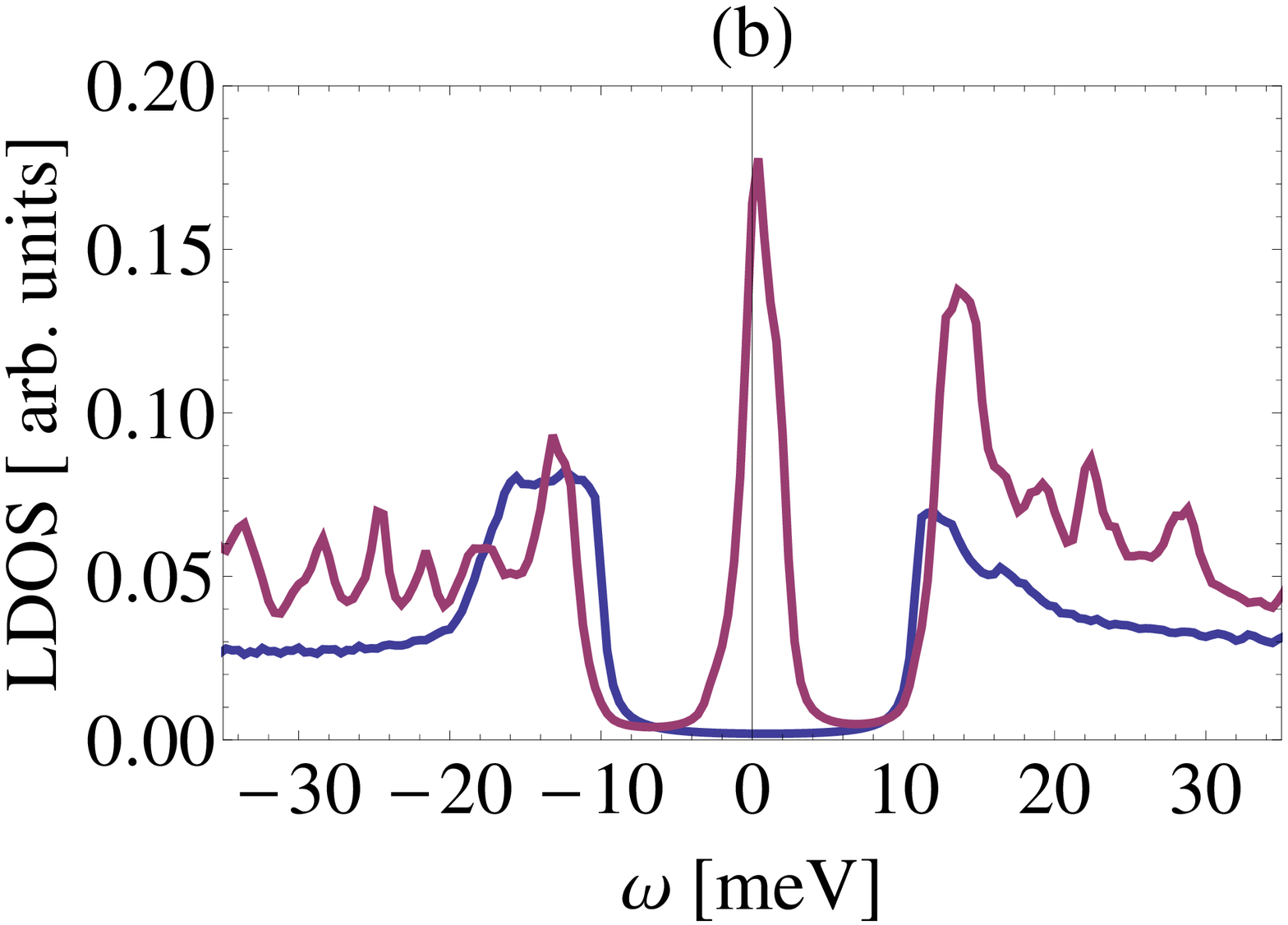}
\end{minipage}
\caption{(Color online) (a) Real-space map showing local magnetization induced by a nonmagnetic impurity potential in the $d$-wave superconducting phase. (b) LDOS for homogenous (blue) and nearest neighbor point (purple) for a nonmagnetic impurity potential. Here impurity potential $V_{imp}=4$eV and Coulomb interaction $U=0.95$eV and $J=U/4$. The electron density is fixed at $n=6.19$ electrons/Fe. } \label{fig:maginduced}
\end{figure}

\section{nonmagnetic impurity-induced magnetism}

In the presence of nonmagnetic impurities in a $d$-wave superconductor, local magnetic order may be pinned around the impurity site if the Coulomb interaction is strong enough to cause a local Stoner instability and not large enough to cross the bulk Stoner criterion (which is close to $U\sim1$eV, $J=0.25U$ in the present case).\cite{tsuchiura01,wang02,zhu02,chen04,kontani06,harter07,andersen07,alloul09,andersen10,christensen11,roemer12,gastiasoro13,gastiasoroprl13} Impurity-induced magnetism is found to be absent for $s$-wave superconductors in agreement with the fact that they do not support sub-gap bound states and hence cannot produce the large DOS enhancement required for locally crossing the Stoner instability. We have calculated the local induced moments and associated LDOS for a $d$-wave superconductor shown in Fig.~\ref{fig:maginduced}(a,b) using a BdG formalism similar to Refs.~\onlinecite{gastiasoro13,gastiasoroprl13,Mukherjee:2013} for the case with $U=0.95$eV and $J=U/4$ and nonmagnetic impurity potential of $V_{imp}=4$eV. The average particle density was kept fixed at $n=6.19$ electrons/Fe. In Fig.~\ref{fig:maginduced}(a) we show the magnetization generated around an impurity. Owing to the dominant $(\pi,\pi)$ nesting over the Fermi surface, we find that the impurity induced moment also forms a local $(\pi,\pi)$ magnetic order. This leads to a bound state peak in the LDOS close to zero bias as seen from Fig.~\ref{fig:maginduced}(b). Note that in the usual case of a nonmagnetic impurity with $V_{imp}=4$eV the LDOS bound state peak is expected to lie close to the gap edge as seen from Fig.~\ref{fig:realgreen}. Hence the contribution to the peak position is primarily from the induced magnetic moments. Thus the presence of induced magnetic order shifts the sub-gap peak positions in the case of $d$-wave superconductivity. However, the fact that nonmagnetic disorder in $s$-wave superconducting states generates neither sub-gap states nor induced order (which would generate effective magnetic impurities and hence sub-gap states) means that the existence of sub-gap bound state is a robust distinctive feature between $d$- and $s$-wave superconductivity in the alkali doped iron chalcogenide superconductors.

\section{Conclusions}

The presence of attractive or repulsive nonmagnetic impurities in the alkali doped iron chalcogenide superconductor K$_x$Fe$_{2-y}$Se$_2$ leads to sub-gap bound states for $d$-wave pairing symmetry, and does not produce any such states for the $s$-wave superconducting state. By studying the effect of changes in chemical potential we have shown that the bound state positions move towards zero bias upon electron doping. This result suggests that it may be favorable to search for sub-gap conductance peaks in samples with larger electron doping since there they are likely to be easier to resolve from the gap edge. Additionally, in the presence of magnetic impurities we find that multiple sub-gap peaks are generated for both $d$-wave and $s$-wave superconducting states. This is in agreement with the presence of sub-gap bound state peaks near interface between $s$-wave or $d$-wave superconductors and a vacancy ordered BAFM state in calculations performed recently in Ref.~\onlinecite{Mukherjee:2013}. Lastly, we have discovered the existence of impurity-induced magnetism generated by (bare) nonmagnetic scattering potentials in the case of $d$-wave pairing. Since no such effect exists for the $s$-wave case, the presence of sub-gap peaks as a measure for the pairing symmetry remains robust for these materials.

B.M.A. and M.N.G. acknowledge support from the Lundbeckfond fellowship (grant A9318).

\end{document}